\documentclass[aps,prl,reprint,superscriptaddress]{revtex4-1}

\usepackage{graphicx}
\usepackage{braket}
\usepackage{amsmath,amssymb,amsfonts}
\usepackage{hyperref}
\usepackage{enumitem}

\usepackage{xcolor}
\usepackage{soul,xcolor}

\newcommand{\fref}[1]{Fig.~\ref{#1}}

\begin{document}

\title
{
The quantum solitons  atomtronic interference device
}

\author{Juan Polo} 
\affiliation{Quantum Systems Unit, Okinawa Institute of Science and Technology Graduate University, Onna, Okinawa 904-0495, Japan}
\author {Piero Naldesi} 
\affiliation{Center  for  Quantum  Physics,  University  of  Innsbruck,  Innsbruck,  Austria}
\author{Anna Minguzzi}
\affiliation{Univ.~Grenoble-Alpes, LPMMC, F-38000 Grenoble, France and CNRS, LPMMC, F-38000 Grenoble, France}

\author{Luigi Amico}

\altaffiliation{On leave from Dipartimento di Fisica e Astronomia 'Ettore Majorana', Universit\`a di Catania, Italy}
\affiliation{Quantum Research Centre, Technology Innovation Institute, Abu Dhabi, UAE}
\affiliation{Centre for Quantum Technologies, National University of Singapore,
3 Science Drive 2, Singapore 117543, Singapore}
\affiliation{CNR-MATIS-IMM \&   INFN-Sezione di Catania, Via S. Sofia 64, 95127 Catania, Italy}
\affiliation{LANEF {\it 'Chaire d'excellence'}, Universit\`e Grenoble-Alpes \& CNRS, F-38000 Grenoble, France}

%%%%%%%%%%%%%%%%%%%%%%%%%%%%%%%%%%%%%%%%%%%%%%%%%%%%

\begin{abstract}
We study a quantum many-body system of attracting bosons confined in a ring-shaped potential and interrupted by a weak link. With such architecture, the system  defines  atomtronic quantum interference  devices harnessing quantum solitonic currents. We demonstrate that the system is characterized by the specific interplay between the interaction and the strength of the weak link. In particular, we find that, depending on the operating conditions,  the current can be a universal function  of the relative size between the strength of the impurity and interaction. The low lying many-body states are studied through a quench dynamical protocol that is the atomtronic counterpart of Rabi interferometry. With this approach, we demonstrate how our system defines a two level system of coupled solitonic currents. 
%{\it qubit}.
The current states are addressed through the analysis of the momentum distribution.
\end{abstract}

\maketitle
{\it Introduction.}
In quantum technology, the boundary between basic and applied research is particularly blurred\cite{dowling2003quantum}. Indeed, interesting quantum devices for applications in this discipline must be constructed with physical platforms characterized by pronounced quantum effects.  At the same time, because of  the specific operating  conditions in which  it  needs to work, quantum  matter designed for quantum technology may display  new fundamental and   unexpected physical features. 

Atomtronics is an emergent field of quantum technology in which the logic described above works as the  core principle.  Such a field  aims at devising a new type of circuitry with degenerate atomic currents both to fabricate devices of practical values and to address basic science\cite{Amico_Atomtronics,amico2020roadmap}.   Key features of the atomtronic platform are the charge-neutrality and coherence properties of the fluid flowing in the circuits that may have fermionic/bosonic nature,  the enhanced control and the versatility of the operating conditions of the circuit elements. 
A fruitful starting point in the current research has been considering ultra-cold matter-wave analogues of known electronic or quantum electronic systems\cite{seaman2007atomtronics,pepino2009atomtronic}.  In particular, ring-shaped  condensates  interrupted  by  one or several weak  links and pierced by an effective magnetic flux\cite{dalibard2011colloquium} have been studied  in analogy with the SQUIDs of  mesoscopic superconductivity\cite{wright2013driving,ramanathan2011superflow,ryu2013experimental,PhysRevA.84.053604,yakimenko2015vortices,yakimenko2015vortices,Turpin:15,mathey2014decay,mathey2016realizing}.  
Such systems, dubbed  Atomtronics  Quantum
Interference Devices (AQUIDs),  enclose a great potential both for basic science and technology\cite{eckel2014hysteresis,ryu2013experimental,hallwood2006macroscopic,solenov2010metastable,amico2014superfluid,aghamalyan2015coherent,aghamalyan2016atomtronic,cominotti2014optimal,ryu2020quantum}
. 

AQUIDs studied so far have been focused on bosons with repulsive interactions.
Atomtronic platforms, though, allow access to much wider physical scenarios that are difficult, if not impossible, to consider otherwise.  
For instance, bosonic mixtures with attractive and repulsive inter- intra-species interactions  \cite{richaud2019pathway}, systems with losses \cite{naether2015stationary} as well as currents in open Bose-Hubbard quantum system have been studied \cite{bychek2020open}
Here, we work with a  quantum many-body system of bosons with {\it attractive} interaction. In one spatial dimension, such a many-body system, recently demonstrated  to be sustaining the quantum analog of bright solitons\cite{naldesi2019rise,polo2019exact}, potentially leads to an  enhancement of sensitivity  to variation  of effective magnetic flux beyond the quantum standard limit \cite{naldesi2019angular}. 
At the same time, one dimensional (1D) attracting bosons with a localized impurity  are very interesting  many-body systems. In this context, the seminal work of Kane and Fisher on fermionic Luttinger liquids defines a true paradigm for the physics of the system. Accordingly, in the  renormalization group sense,  a single localized impurity in an attractive (repulsive) fermionic system  is suppressed (enhanced)   by the interaction\cite{kane1992transport}. 
Our system does not fall in the  Kane-Fisher scheme for several reasons. 
First, our system is made of attracting bosons.  In fact, despite low energy attractive fermions  being expected to behave as repulsive bosons \cite{cazalilla2004bosonizing,cazalilla2011one,Krinner_2017}, attractive bosons display a quadratic dispersion\cite{calabrese2007correlation}.    Recent studies  do indicate that strongly attractive bosons (super Tonks regime) can  define a Luttinger liquid, but such a state is a specific excited state \cite{astrakharchik2005beyond,batchelor2005evidence}. Second, our set up  is of  mesoscopic size. The persistent current for repulsive bosons  on a mesoscopic-size ring was studied recently\cite{cominotti2014optimal}. Particularly, the impurity results to be  suppressed by interaction, but to a finite value, with  the persistent current  displaying a  remarkable non-monotonous dependence on interaction.  

We note that at intermediate interactions the  mesoscopic regimes of  attractive bosons in presence of a localized  impurity are not known. 

In this paper, we harness the features of an attracting many-boson quantum fluid to define the atomtronic device based on entangled solitonic currents: The quantum Solitons Atomtronic Quantum Interference Device (S-AQUID).  In our system, the quantum fluid flows in a mesoscopic ring-shaped  potential and interrupted by a single weak link. We will show that our solitonic current has peculiar transmission properties.  While we find that the interaction can pin the quantum soliton,  the actual interplay between transmission and interaction substantially departs from Luttinger liquid behaviour. 
The interplay of such parameters is important for the generation of specific states of entangled solitonic currents. In certain regimes, we demonstrate that the system is characterized by a two-level system dynamics.
For the analysis of such states, we  devise a specific quench protocol defining the  atomtronic counterpart of the  Rabi-type measurement protocol of the persistent current.     
Finally, we show that the  read-out of the system 
can be carried out by a specific analysis of the atomic cloud after the free expansion of the system.

{\it Model system}.
We consider a system of $N$ interacting bosons 
loaded into a 1D ring-shaped optical lattice of $N_s$ sites.
The discrete rotational symmetry of the lattice ring is broken by the presence
of a localized potential on one lattice site, which gives rise to a weak link.
The ring is pierced by an artificial  magnetic flux $\Omega$.
In the tight-binding approximation, this system is described
by the 1D Bose-Hubbard (BH) Hamiltonian
\begin{equation}
\hat{\mathcal{H}}(\Omega)= \sum_{j=1}^{N_s} \left[\frac{U}{2} n_{j}\left( {n}_{j} \!-\! 1\right) \!-\! J\left( e^{-i \tilde{\Omega}}a_{j}^{\dagger }a_{j+1} \!+\! \text{h.c.} 
\right) \!+\! \lambda_{j} n_{j}\right] ,
\label{BHH}
\end{equation}
where $a_{j}$ and $a_{j}^{\dagger }$ are site $j$ annihilation and creation Bose operators and $n_{j}\!=\!a_{j}^{\dagger }a_{j}$. The parameters $J$ and $U$ in \eqref{BHH} are respectively the hopping amplitude and the strength of the on-site interaction. Here we consider $U<0$ to describe the particles attraction.  The presence of the flux $\Omega$ is taken into account through the Peierls substitution:
$J\rightarrow J e^{-i\tilde{\Omega}}$ with  $\tilde{\Omega}\doteq 2\pi \Omega/(\Omega_0N_s)$, with $\Omega_0$ the single-particle flux quantum. The potential barrier considered here is localized on a single site $j_0$,
i.e., $\lambda_{j}=\lambda\delta_{j,j_0}$ with $\delta_{i,j}$ being the Kronecker delta.

In the absence of the barrier $\lambda=0$, the  ground state of the system (\ref{BHH})  is a bound state of solitonic nature with  approximately quadratic dispersion, becoming flat as $|U|N$ increases. For any finite negative interaction, extended (or scattering) states are separated from the bound state by a finite energy gap increasing with 
$|U|$ (see detailed definitions and characteristics of these states in \cite{naldesi2019rise}).  For sufficiently large $|U|$, any bound state is separated from (the band of) extended states by a finite energy gap. The latter  feature is a genuine lattice effect\cite{naldesi2019rise,polo2019exact}. 
The ground state energies  for different $\Omega$ displays specific degeneracies with  a periodicity, fixed by an elementary flux quantum $\Omega_p$, that depends on the number of particles and 
on the interaction \cite{naldesi2019angular}. This feature provides a generalization of the Byers-Yang pairing states\cite{byers1961theoretical} and the Leggett theorem \cite{nanoelectronics1991dk}.   

In the dilute limit of small filling fractions $N/N_s \ll 1$, the BHM is equivalent to the Lieb-Liniger model of bosons with a delta localized barrier.
The Lieb Liniger model (with no barrier) is  exactly solvable by Bethe Ansatz. In particular, the dispersion relation for the low lying excitations is quadratic in the wave vector: $\omega = \hbar k^2/ mN$ and becomes flat in the limit of a large number of particles\cite{calabrese2007correlation}. The character of the field theory that can describe the low-energy excitations on top of the ground state  
remains unclear \cite{astrakharchik2005beyond,batchelor2005evidence}. The super-Tonks regime was proved to be obtained as a highly excited state of the Lieb-Liniger model\cite{batchelor2005evidence}; excitations on top of such a state can be described with a Luttinger liquid theory.  Similarly to the BHM (\ref{BHH}), the ground state of the Lieb-Liniger model is characterized by $1/N$-fractionalization of $\Omega_0$.  

\begin{figure}
\includegraphics[width=\columnwidth ]{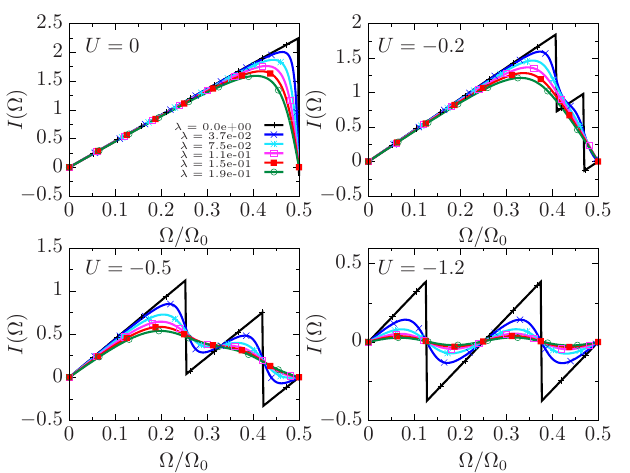}
\put(-245,180){(a)}
\put(-127,180){(b)}
\put(-245,90){(c)}
\put(-127,90){(d)}
\caption{{\it The Persistent current}. Current as a function of  $\Omega/\Omega_0$ for different values of the barrier strength. (a) Currents show the smoothening of the sawtooth behavior at $U=0$ as the barrier, $\lambda$, increases. (b-c) Shows the transition from the single periodicity of the current to an $N$-times periodicity. Strong barriers can remove the sawtooth spikes of the current. (d) Full $N$-times increase of the periodicity of the current. Strong barriers smoothen the current, but the periodicity is preserved.
The results are obtained with exact diagonalization with   $N=4$ and $N_s=11$.}
\label{barrierVSinteraction}
\end{figure}
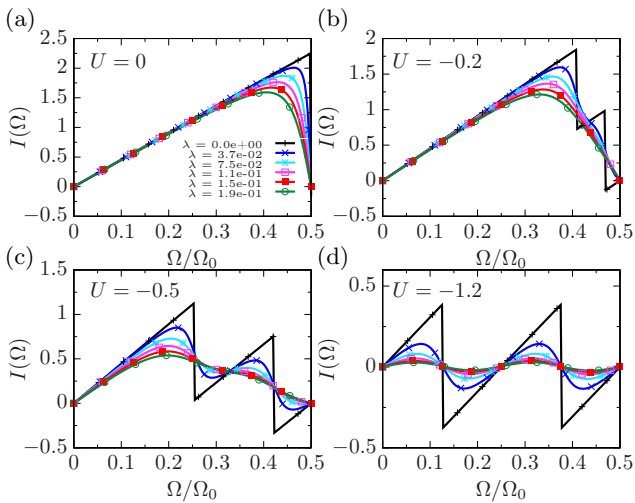

In this paper, we monitor the  ground state persistent current $ 
\label{eq:2}
I(\Omega ) = -{\partial E_{0}}/{\partial\Omega}
$
where $E_0$ is the ground state energy. For the model (\ref{BHH}), 
$ I(\Omega )$ is given by 
\begin{equation}
\smash{I(\Omega)=-i J\sum_{j}\langle e^{-i \tilde{\Omega}}a^{\dagger}_{j}a_{j+1} - e^{+i \tilde{\Omega}}a^{\dagger}_{j+1}a_{j}\rangle_0}
\end{equation} 
where $\langle \bullet \rangle_0$ is the groundstate expectation value.
For a  quantum system in a ring,  the  angular momentum is quantized~(see \cite{moulder2012quantized,wright2013driving} for recent experiments).  Accordingly, $I(\Omega )$ displays a characteristic  sawtooth behaviour, with a periodicity that Leggett  proved to be fixed by the elementary  flux quantum of the system~\cite{byers1961theoretical,onsager1961magnetic,nanoelectronics1991dk}. For repelling bosons the latter quantity is $\Omega_0$; for attractive interactions instead the elementary flux quantum can get fractional values $\Omega_0/N$- see \fref{barrierVSinteraction}. In other words, the ground state of attracting bosons will present {\it current states} at fractional values of $\Omega_0$, ie states with non-vanishing persistent current. 

{\it Interplay between barrier and interaction}. 
Below, we study the configuration of energy levels and persistent currents as function of  interaction and barrier strength. 
In the absence of impurity, the ground state results to be degenerate at specific values of $\Omega$.
For $\lambda \neq 0$, the  degeneracies are lifted - see Fig. \ref{fig:TOF}(b) with  hybridized ground and low lying scattering states. For large $\lambda/|U|$, the state is characterized by a  substantial overlap with the scattering states and therefore, the physics is expected to be governed by  latter ones.  On the other hand, for small $\lambda/|U|$, the state is nearly localized with a small weight of the scattering states. In this regime, the solitonic nature of the matter-wave plays a significant role. 
The persistent current is affected dramatically by the interplay described above -  Fig.\ref{barrierVSinteraction}. For small  $\lambda/|U|$, the  weak link is not able to break the bound state and the resulting solitonic nature of the current suppresses the transmission through the barrier; accordingly, the persistent current displays  oscillations with  a reduced  period  reflecting the fractionalization of $\Omega_0$. 
For larger  $\lambda/|U|$, instead, the matter wave can be split in transmitted/reflected amplitudes and therefore the current is not fractionalized, displaying  the same periodicity $\Omega_0$ as the  repulsive bosons cases. 

\begin{figure}
\includegraphics[width=1\columnwidth]{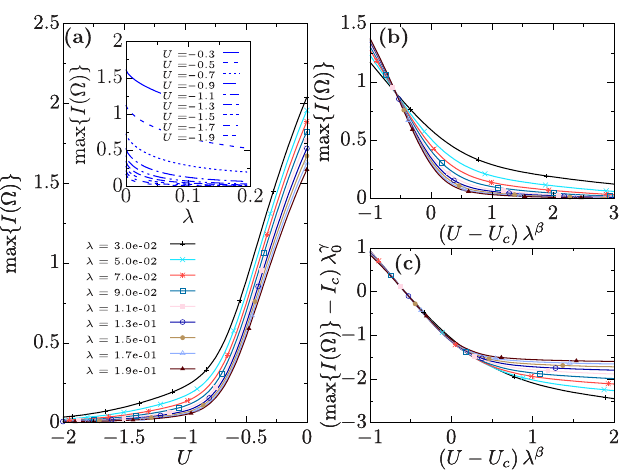}
\caption{
{\it Universal behavior of the persistent current}. (a) Maximum of persistent current as a function of the interactions $U$. Inset shows $\textrm{max}\{I\}$ versus the barrier strength $\lambda$. (b) Maximum of the persistent current as a function of rescaled interactions for the different barrier strengths $\lambda$: $\textrm{max}\{I\} = g\left((U-U_c)\times\lambda^{\beta}\right)$.
(c) Collapse of the behavior of the maximum of the persistent current: $(\textrm{max}\{I\}-I_c)\times \lambda^{\gamma}  = g\left((U-U_c)\times\lambda^{\beta}\right)$. Parameter values are: $\beta = -0.4$, $U_c =  0.633$, $\gamma = -0.35$ and $I_c =  0.9025$.
Particle number and number of sites are as in Fig.\ref{barrierVSinteraction}.
}
\label{fig:collapse}
\end{figure}

\begin{figure*}[h!!!!t!!!]
\includegraphics[width=2\columnwidth]{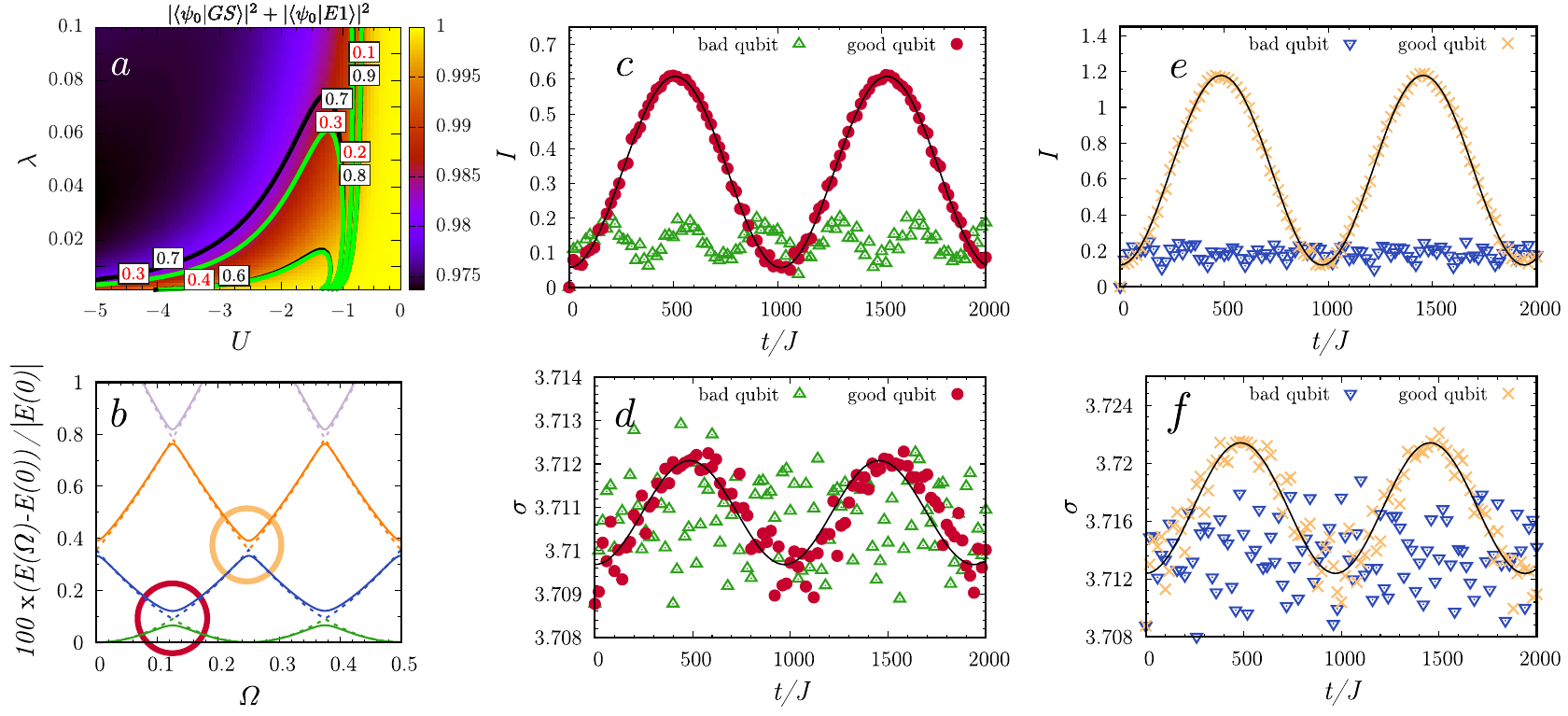}
\caption{{\it Rabi measurement protocol of the S-AQUID}.  (\textbf{a}) Projection, $|\langle \psi_0 | GS \rangle|^2 + |\langle \psi_0 | E1 \rangle|^2$, after quenching  $\Omega/\Omega_0$ from $0$ to $0.125$ with $\ket{GS}$ being the ground state and $\ket{E1}$ the first excited states. The solid lines indicate  'equipotential lines' of constant $|\langle \psi_0 | GS \rangle|^2$ (black-solid lines) and $|\langle \psi_0 | E1 \rangle|^2$ (green-solid lines). The actual constant values are indicated in the white boxes, with red text corresponding to green curves and black text to black lines. (\textbf{b}) 
shows the energy spectrum of the model Eq.(\ref{BHH}) as a function of flux without barrier (dashed lines) and for barrier $\lambda=0.01$. Oscillations of the persistent current $I$, panels (\textbf{c}) and (\textbf{e}), in the laboratory frame and width of the momentum distribution $\sigma$, panels (\textbf{d}) and (\textbf{f}), as a function of time. 
The two states are obtained by quenching the value of the flux to the two corresponding crossing points, marked in the  panel (\textbf{b}) in red and yellow respectively.
In (\textbf{c}) and (\textbf{d}) the good TLS is created by quenching from $\Omega_i/\Omega_0=0$ to  $\Omega_f/\Omega_0=0.125$ while the bad one is created by quenching to $\Omega_f/\Omega_0=0.175$. In (\textbf{e}) and (\textbf{f})  the good TLS is created by quenching from $\Omega_i/\Omega_0=0$ to  $\Omega_f/\Omega_0=0.25$ while the bad one is created quenching to $\Omega_f/\Omega_0=0.3$. Interaction and barrier are fixed to $\lambda=0.01$ and $U=-1.5$. Number of particles and lattices sites are as in Fig.\ref{barrierVSinteraction} and  Fig.\ref{fig:collapse}. All numerical results are obtained through exact diagonalizaiton (see Supplemental).}
\label{fig:TOF}
\end{figure*}

We analyse the interplay beteween the barrier and interactions by monitoring the persistent current amplitude i.e. $ \max\{I(\Omega)\}$   
-  see Fig.\ref{fig:collapse}. 
We identify the emergence of  two regimes, separated by the transition of the current from a single sawtooth to the $N$-times periodicity. At increasing
interactions, the maximum of the current decreases and smaller sawtooths start to appear in the interval $\Omega/\Omega_0 = [0,1/2]$.  In this regime the current is found to be a  function of $\lambda/|U|$, with a clear data collapse shown in Fig.~\ref{fig:collapse}(c), indicating a non-trivial interplay between interaction and barrier strengths \cite{barber1983phase}. We note that for $N=\{5,\,6\}$ the collapse  occurs with unaltered exponents $\beta$ and $\gamma$; while $I_c$ and $U_c$ display a weak $N$-dependence (see Supplemental). 
On the other hand, for large enough interactions a $N$-times periodicity is reached and the persistent current amplitude depends on $U$ and $\lambda$ separately, marking the  break-down of the collapse.   Finally, we also observe that at  
large barriers the energy band dispersion   flattens and the gaps between energy bands increase,
%${U,\lambda}$
yielding again a breakdown of the collapse.

{\it Rabi oscillations}.
To probe entangled current states, we adopt a specific quench  protocol providing the atomtronic counterpart of Rabi spectroscopy\cite{1950PhRv...78..695R}:
We start  the system in a state with  zero persistent current  $|\psi_0 \rangle \doteq|\psi(\Omega_i\!=\!0) \rangle$ and perform a quench on the effective magnetic flux to the final value $\Omega_f$: $|\psi (t) \rangle=\exp[-i\hbar \hat{\mathcal{H}}(\Omega_f) t] |\psi_0) \rangle$.  This way, the $|\psi (t) \rangle$-expectation value of the current $I(\Omega_f)$ will display characteristic oscillations in time, corresponding  to superposition of different current states (see \fref{fig:TOF} (c-d)).

In \fref{fig:TOF}(a), we systematically study how the  projection and the amplitude of each eigenstate after the quench depends on interaction and barrier strength. Note that, although the projection between the initial state and the ground state plus first excited states of the post-quenched system is always large, only for a specific  parameter regime 
%a 
ideal
two-level system (TLS) dynamics arises,  with equal projection into each states (solid lines in \fref{fig:TOF}(a) with values $(0.5,0.5)$ indicating a balanced population of the two states after quenching).  
%More specifically, we study how  a %\blue{TLS emerges in  the energy %landscape (see Fig.\ref{fig:TOF}(b)) %with no other excited states %involved in the dynamics. 
Our TLS is of the form
$\frac{1}{2}\left( \ket{GS} + e^{i\theta}\ket{E1} \right)$, $\theta$ being an initial relative phase, describing coupled solitonic currents.
The quench induces Rabi oscillations in the current with frequency pattern reflecting the actual many-body spectrum  of the system.
While oscillations with multiple frequencies occur for  generic values of the  parameters, \fref{fig:TOF} (green/blue-triangles), we observe that suitable combinations of $U$, $\lambda$ and $\Omega$ lead  to Rabi oscillations with a single frequency corresponding to transitions between just two energy levels: The TLS dynamics.  
We note that, because of the specific features of the attractive boson interaction, the protocol  allows us creating TLS at fractional values of $\Omega_0$ $\Omega=\Omega_0/2N$ - \fref{fig:TOF} (red circles); this feature should be contrasted with the standard  AQUID (repulsive bosons) in which the degeneracy points occur at odd-integer multiples of $\Omega_0/2$. Therefore,  following the same protocol in the regime where the S-AQUID presents $N$-times periodicity of the current, our protocol can  entangle current states with distant angular momenta. That is, by quenching from the ground state at $\Omega_i=0$ to, for instance, a rotation of $\Omega_f=2\times\Omega_0/2N$ we can create a TLS of the form $\frac{1}{2}\left( \ket{E1} + e^{i\theta}\ket{E2} \right)$ - see \fref{fig:TOF}(e-f).

{\it Readout}. Despite the peculiar coherence properties of the solitonic ground state, the momentum distribution can  characterize the pattern of currents flowing in the system. Specifically, non vanishing and quantized currents are detected by the width $\sigma$ of momentum  distribution\cite{naldesi2019angular}. 
Here we demonstrate that the time evolution of $\sigma$ after the quench can be used to monitor the quantum dynamics of the system. Specifically, the  width $\sigma(t)$ of the momentum distribution is
\begin{eqnarray}
\nonumber
\sigma(t) = \sqrt{ \int d{\bf k} \: {\bf k}^2 \: n({\bf k},t) } \; ;
\end{eqnarray}
where $ 
n({\bf k},t) = \sum_{i,j} \: e^{i\bf k\cdot ({\bf R}_i-{\bf R}_j)}  C_{i,j}(t) 
$, 
with $C_{i,j}(t) = \langle \psi(t) | a^{\dagger}_ia_j | \psi(t) \rangle$.  The results are shown in \fref{fig:TOF} d,f. We note that, while in the TLS regime $\sigma(t)$ displays Rabi oscillations with the current periodicity,   multiple  frequencies emerge in its dynamics for the  physical regimes in which  many states contribute in the current superposition. 

{\it Conclusions}. We have studied an electrically neutral quantum fluid of attracting bosons confined in a ring-shape potential of mesoscopic size, interrupted by a localized tunnel barrier and pierced  by an effective magnetic field. 
 We point out that the peculiar interplay between number of particles  and interaction characterizing our system    \cite{calabrese2007correlation,naldesi2019rise}, makes our approach (small $N$ and finite $U$) especially relevant.
 
 Because of  the quantum solitonic nature of the ground state and  its mesoscopic size the  system defines a quantum fluid with unique features. Indeed, the transmission through the barrier is dramatically affected by the interaction (see  Fig.\ref{barrierVSinteraction}). The physics departs from the Luttinger liquid paradigm for which an arbitrary small impurity should be able to pin the soliton. Indeed, the specific pinning features of the quantum soliton imply that the  persistent current is characterized by an interplay between the impurity strength and interaction that is found to display universal features only in specific regimes (see Fig.\ref{fig:collapse}). 

Our system provides a matter-wave circuit that is  the solitonic counterpart of the atomic SQUID: The S-AQUID. Due to the peculiar coherence of the quantum fluid hardware, the S-AQUID is characterized by specific physical properties, implying, in turn, unique features. 
In particular, we point out that, in contrast with the standard implementations exploiting repulsive bosons, the TLS dynamics emerge at fractional values of the elementary flux quantum $\Omega_0$. 
% We have shown that our system can define a qubit. 
%
In analogy with SQUIDs, such TLS, macroscopic superposition of quantum solitons, can be relevant for quantum sensing \cite{naldesi2019angular}
To address the  quantum-coherent dynamics,  we devised  the atomtronic counterpart of the Rabi measurement protocol. 
We demonstrated that the quench dynamics of the system can be read-out  by a specific analysis of the momentum distribution - Fig.~\ref{fig:TOF}.  Most of our results are within  the current know-how in cold atoms quantum technology, and are particularly relevant on ring geometries \cite{deherve2021versatile,pandey2021atomtronic,henderson2009experimental,lesanovsky2007timeaverage}. 

We point out that in the lab frame, 
Fig.\ref{barrierVSinteraction} produces a staircase dependence of the current, with each plateau corresponding to a quantized value,
on the scale  $\Omega_p=\Omega_0/N$ (see for instance \cite{naldesi2019angular}).
Being then the effective magnetic field related to a specific current quantum number,
 our system can perform an {\it absolute} measurement (after calibration), with resolution 
fixed by the  fractional  magnetic flux. 
This 
should be contrasted with the standard 
SQUIDs or single electron transistors protocols 
performing
differential measurements at fixed quantized value of the current
\cite{devoret2000amplifying,tinkham2004introduction}. By realizing the magnetic field by a rotation \cite{dalibard2011colloquium}, the S-AQUIDs open the way to a rotation sensing device with enhanced ($1/N$) sensitivity, approaching the Heisenberg limit in  atomic interferometry \cite{naldesi2019angular}.

{\it Acknowledgements}. It is a pleasure to thank Gianluigi Catelani, Maxim Olshanii  and Helene Perrin for discussions. The Grenoble LANEF framework ANR-10-LABX-51-01 are acknowledged for their support with mutualized infrastructure. JP acknowledges funding from the JSPS KAKENHI Grant No. 20K14417. 

\bibliographystyle{apsrev4-1}

\bibliography{references.bib,references_2.bib,library.bib}

%%%%%%%%%%%%%%%%%%%%%%%%%%%%%%%%%%%%%%%%%%%%%%%%%%%%%%%%%%%%%%%%%%%%%%
%%%%%%%%%%%%%%%%%%%%%%%%%%%%%%%%%%%%%%%%%%%%%%%%%%%%%%%%%%%%%%%%%%%%%%
%%%%%%%%%%%%%%%%%%%%%%%%%%%%%%%%%%%%%%%%%%%%%%%%%%%%%%%%%%%%%%%%%%%%%%%%%%%%%%%%%%%%%%%%%%%%%%%%%%%%%%%%%%%%%%%%%%%%%%%%%%%%%%%%%%%%%%%%%%%%%%%%%%%%%%%%%%%%%%%%%%%%%%%%%%%%%%%%%%%%%%%%%%%%%%%%%%%%%%%%%%%%%%%%%%%%%%%%%%%%%%%%%%%%%%%%%%%%%%%%%%%%%%%%%%%%%%%%%%%%%%%%%%%%%%%%%%%%%%%%%%%%%%%%%%%%%%%%%%%%%%%%%%%%%%%%%%%%%%%%%%%%%%%%%%%%%%%%%%%%%%%%%%%%%%%%%%%%%%%%%%%%%%%%%%%%%%%%%%%%%%%%%%%%%%%%%%%%%%%%%%%%%%%%%%%%%%%%%%%%%%%%%%%%%%%%%%%%%%%%%%%%%%%%%%%%%%%%%%%%%%%%%%%%%%%%%%%%%%%%%%%%%%%%%%

\end{document}